\documentclass[preprint,showpacs,preprintnumbers,amsmath,amssymb]{revtex4}
 \usepackage{amsfonts}
\usepackage{amsmath}
\usepackage{amssymb}
\usepackage{graphicx}%
\begin{document}
\title{Transition from KPZ to Tilted Interface Critical Behavior in a Solvable
Asymmetric Avalanche Model}
\author{A.M. Povolotsky$^{1,2}$ }
\author{V.B. Priezzhev$^{2}$}
\author{Chin-Kun Hu$^{1}$}
\email{huck@phys.sinica.edu.tw} 
\affiliation{$^{1}$Institute of Physics, Academia Sinica, Nankang, Taipei 11529, Taiwan}
\affiliation{$^{2}$Bogoliubov Laboratory of Theoretical Physics, J.I.N.R., Dubna 141980, Russia}
\begin{abstract}
We use a discrete-time formulation to study the asymmetric avalanche process [Phys. Rev. Lett.
\textbf{87}, 084301 (2001)] on a finite ring and obtain  an exact
expression for the average avalanche size of particles as a function of
toppling probabilities depending on parameters $\mu$ and $\alpha$. 
By  mapping the model below  and above the critical line onto driven 
interface problems, we  show how different regimes of avalanches may lead to different
types of critical interface behavior characterized by either
annealed or quenched disorders and obtain exactly the
related critical exponents which violate a well-known scaling
relation when  $\alpha \ne 2$.
\end{abstract}
\maketitle
The physics of driven interfaces has been attracting great attention
for many years due to its connection with a variety of phenomena
such as fluid flow through porous media,
motion of charge density waves and flux lines in superconductors, etc.
\cite{hhz,k,f}. The results obtained can be roughly splitted into two groups.
The first one incorporates growth phenomena where thermal or
annealed fluctuations affect the dynamics. The most famous example
is the Kardar-Parizi-Zhang (KPZ) universality class characterized
by the roughness exponent $\chi=1/2$ and the dynamical exponent
$z=3/2$ in $1+1$ dimensions \cite{KPZ}. In the second group, an
interface moves under the action of an external driving force $F$,
which competes with pinning forces due to quenched medium
inhomogeneities. At some critical force $F_{c}$ the transition
occurs from the totally pinned state to the state where the
interface moves with a velocity $v$, which shows a power law decay
$v\sim\left(  F-F_{c}\right)  ^{\theta}$ when $F$ approaches the
critical point $F_{c}$ from above \cite{NF} .

The scaling properties of interfaces in the quenched case strongly
depend on the medium isotropy. The isotropic medium produces a
rough interface, which is believed to obey the quenched KPZ
equation with a nonlinear term vanishing at the depinning threshold,
$F\rightarrow F_{c}$ \cite{ABS,TKD}. In the anisotropic medium, the
quenched KPZ equation holds only for a definite orientation (hard
direction) of the interface, yielding a divergent nonlinear term.
The tilt from the hard direction generates the gradient term in
the equation which breaks the translation invariance. The
existence of such a term gives rise to another universality class,
tilted interface (TI) class , characterized in 1+1 dimensions by the exact
exponents $\chi=1/2$, $z=1$, and$\ \theta=1$ \cite{ABS,TKD}.

It has been noted that thermal fluctuations below the depinning
threshold can initiate avalanches, which lead to infinitely slow
creep of the interface \cite{tl}. Far below depinning
threshold, the avalanches lead to local advances of finite
interface segments, so that they can be considered under
coarsening as a local annealed disorder in an appropriate time
scale. Approaching the depinning threshold, the local avalanches
increase up to the system size and contribute to a global
interface depinning which is controlled by the quenched disorder.
It is the aim of this Letter to treat the effects of annealed and
quenched disorder in a frame of a unified model.

Due to the well known correspondence between growth processes in
$(1+1)$ dimensions and one-dimensional lattice gases \cite{hhz}, we
can interpret the avalanche dynamics of interfaces in terms of
avalanches of particles similar to those appearing in
one-dimensional random sandpile models \cite{MZ}. To this aim, we
can use the Asymmetric Avalanche Process (ASAP) which has been
recently proposed and solved by the  Bethe ansatz method
\cite{piph}. Having formal origin in the asymmetric exclusion
process \cite{derrida}, the ASAP has completely different
dynamical behavior specific for systems with avalanche dynamics
which has been widely investigated in the context of
self-organized criticality \cite{btw,Dickman}.

The ASAP is formulated as a system of particles diffusing on the
lattice. The interaction makes particles to accelerate if they are
in the same site of the lattice, giving rise to avalanche-like
relaxations of domains with higher density of particles. Such a
dynamics depends on transition rates and may lead to different
regimes of the particle flow, intermittent or continuous.
In the former case, fast avalanche processes alternate with long
periods of relatively slow diffusive motion and the average
avalanche size of particles is finite being measured in the diffusive
time scale. This average avalanche size diverges in the thermodynamics limit if
the density of particles approaches a critical point \cite{piph} .
The divergency means that the diffusive time-scale is no more
appropriate for the description of the system, because the
avalanche processes become dominating. Internal avalanche time
should be used in this case to quantify the dynamics in the
system.

The study of the intermittent flow phase by the Bethe ansatz on a
lattice of finite size $N$ \cite{pph}  shows that the total
distance traveled by particles has the long time asymptotic
$\left\langle Y_{t}^{2}\right\rangle _{c}\sim tN^{3/2}$, which is
equivalent to finding the KPZ dynamical exponent $z=3/2$
\cite{hhz,KPZ}. It will be clear below that in the phase
of continuous flow, Maslov-Zhang (MZ)
arguments \cite{MZ} based on the mapping of the system onto the
random walk problem show that the model belongs to the TI
universality class at least in the vicinity of the transition
point. In this Letter, we use the equivalence between the driven
interface problem and the ASAP in discrete time to study the
crossover between these two types of disorder. Considering the
ASAP on the lattice, we obtain the exact expression for the
average avalanche size  in two different regimes and
describe scaling properties of the system at the transition point.

Consider the system of $p$ particles on a one-dimensional
lattice of $N$ sites with periodical boundary conditions, i.e. a ring. The
system evolves by discrete time steps $\Delta t<1$ according to
the following rules. \textbf{I. } If all particles occupy
different sites of the lattice, then one randomly chosen particle
moves one step to the right neighboring site  with probability
$\Delta t$ for one time step. \textbf{II.} If there are $n>1$
particles at the same site, referred to as an active site, the
diffusive motion is replaced by the avalanche dynamics: either all
$n$ particles move to the next right site at the next time step with
probability $\mu_{n}$ or $n-1$ particles move one step to the next right
site with probability $1-\mu_{n}$, while one particle remains
unmoved. Thus, we use a random sequential update and separate in
time the avalanche and single particle dynamics. The continuous
time limit, $N\to \infty$ and $\Delta t \to 0$, coincides with the case
considered in \cite{piph}, as only the terms of order $\Delta t$
survive in the master equation.

To map the system to the interface problem, we associate the
occupation number $n_{i}$ on a site $i$ with an interface height
increase $n_{i}=h\left(  i+1\right) -h\left(  i\right)$ so that
the number of particles determines the tilt of the interface. To
combine periodic boundary conditions with the interface tilt we
imply helicoidal boundary condition to the interface by putting
$h\left( i+N\right) =h\left(  i\right)  +p$.

A state of the system  at time $t$ is given  by probabilities of
all possible configurations of particles on the lattice. Such
probabilities obey  the  master equation
\begin{equation}
P_{t+\Delta t}(C)=\sum_{\{C'\}}T(C,C') P_{t} (C')
\label{master}
\end{equation}
where $T(C,C')$ denotes the transition probability from the
configuration $C'$ to $C$ and the summation is over all possible
configurations of particles. Despite the ambiguous explicit form
of Eq.(\ref{master}), the form of its stationary solution is
rather simple and can be obtained by direct solution under an
assumption of stationarity. In the  stationary state, the
probability of a configuration depends only on the number of
particles in the active site rather than on occupation of the
other sites of the lattice. Let ${\cal P}^{\left[  n\right]} $ be
  the probability of any configuration with $n$ particles at the active site.
Then, the stationary    solution of the system (\ref{master}) is
given by the following recursion relations
\begin{equation}
\mathcal{P}^{\left[  n+1\right]  }   =\mathcal{P}^{\left[  n\right]  }\mu_{n}/({1-\mu_{n+1}}),
\label{measure}
\end{equation}
where $n=1,2,\ldots,p-1$, and $\mu_1=\Delta t/p $. The only
undetermined constant here is $\mathcal{P}^{\left[  1\right]  }$,
the probability of stable configurations with no active site,
which is defined from the normalization condition
$\sum_{\{C\}}P(C)=1$.

As  it follows from the Eq.(\ref{measure}),
when $n$ particles leave an active site at some
step of an avalanche, all configurations of the remaining $p-n$
particles at the lattice occur equally likely.
This form of the stationary measure is an analogue of the
equiprobability of all configurations in the exclusion process
\cite{derrida}. Then, we can
proceed in a way similar to that in Ref. \cite{piph} and consider
the expected number $P_{k}\left(  n\right)  $ of events when $n$
particles leave an active site at the $k$-th step of an avalanche.
The number of particles, $n$, outgoing from an active  site at
every time step performs Markov random walk
\begin{equation}
P_{k+1}(n)=\sum_{m}P_{k}(m) w_{m, n}
 \label{rw}
\end{equation}
 with  transition probabilities
\begin{align}
w_{n-1,n}  &  =\left(  \rho-\frac{n-1}{N}\right)  \mu_{n}\nonumber\\
w_{n,n}~~~  &  =\left(  \rho-\frac{n}{N}\right)  \left(  1-\mu_{n+1}\right)
+\left(  1-\left(  \rho-\frac{n}{N}\right)  \right)  \mu_{n}\nonumber\\
w_{n+1,n}  &  =\left(  1-\left(  \rho-\frac{n+1}{N}\right)  \right)  \left(
1-\mu_{n+1}\right) \nonumber
\end{align}
for $n>1$ and
$w_{1,1}=\left(  \rho-\frac{1}{N}\right)  \left(1-\mu_{2}\right)  $,
$w_{2,1}=\left(  1-\left( \rho-\frac{2}{N}\right)  \right) \left(1-\mu_{2}\right)  $,
where $\rho=p/N$.
Equation (\ref{rw}) generalizes
corresponding equation in \cite{piph} explicitly taking into
account the effect of the finite size on the transition
probabilities. This enables us to consider the ASAP both in the
intermittent and continuous flow regimes. Instead of $P_{k}(n)$,
we can look for the total expected number of these events during
the whole avalanche $P(n)=\sum_{k=1}^{\infty}P_{k}(n)$, which
satisfies much simpler equation
$P(n)=\delta_{n,1}+\sum_{m}w_{mn}P(m)$, supplemented by the
initial condition $P\left(  1\right)  \left(
1-\frac{p-1}{N}\right)  =1$. The solution of this equation can be
found from the detailed balance condition, $P\left(  n\right)
w_{n,n+1}=P\left(  n+1\right)  w_{n+1,n}$. The result is
\begin{equation}
P\left(  n\right)  =\frac{N\Gamma(p)\Gamma(N-p+1)}{\Gamma(p-n+1)\Gamma
(N-p+n+1)}\prod\limits_{j=2}^{n}\frac{\mu_{j}}{1-\mu_{j}}, \label{P(n)}%
\end{equation}
for $n>1$.  The expected total number $V$ of particles spilled during an
avalanche, i.e. the average avalanche size,  is given by
\begin{equation}
V=\sum\limits_{n=1}^{p}nP\left(  n\right)  . \label{V}
\end{equation}
As a single-particle jump which can trigger an avalanche occurs with
probability $\Delta t,$ the average number of steps of particles for time
$\Delta t$ is $\Delta tV$, if one neglects the time taken by the avalanche itself
in comparison with the average time interval between avalanches.
Therefore, if $\Delta t\to 0$ and $V$ is finite, we may consider  $V$ as an average
velocity of particles or the velocity of the associated interface in the
continuous diffusive time scale.

Consider the asymptotic behavior of the average avalanche size, $V$,  in the
limit $N\rightarrow\infty,p\rightarrow\infty,p/N=\rho=const$. To
this end, we have to specify the toppling probabilities $\mu_{j}$.
We suppose that $\mu_{j}$ tends to a constant limit
$\mu_{\infty}$  when $j\rightarrow\infty$, and define the function
$f\left(  n\right)  $ through the relation
${\mu_{n}}/({1-\mu_{n}})=e^{f\left( n\right)
}{\mu_{\infty}}/({1-\mu_{\infty}})$, assuming that it has the
expansion $f\left(  n\right)  \sim (\alpha-2)/n+O(1/n^{2})$. It
turns out that $\alpha$ is the only parameter which is responsible
for the subcritical singularity of the average avalanche size. Below we
imply $\alpha$ to vary in the range $1<\alpha<3$.
  In the thermodynamic limit we have  the
leading term of the $n/N$ expansion
$V\sim\sum_{n=1}^{\infty}n^{\alpha-1}\left( \frac
{\mu_{\infty}\rho}{\left(  1-\mu_{\infty}\right)  \left(
1-\rho\right) }\right)  ^{n-1} $, which converges for $\alpha>0$
if the density of particle is less than the
critical value $\rho<\rho_{c}$, where $\rho_{c}=1-\mu_{\infty}$.
Close to the critical point, $V$ diverges with the critical exponent $\alpha$,%
\begin{equation}
V\sim\left(  \rho_{c} -\rho\right)  ^{-\alpha}.
\label{Vsubcrit}
\end{equation}

Above the critical line, we use the Stirling formula to
approximate gamma functions in Eq. (\ref{P(n)}). The leading order
of the expansion shows that the function $P\left(  n\right)  $ has
a maximum at the point
\begin{equation}
n_{0}=N\left(  \rho-\rho_{c}\right)  \text{.} \label{n0}%
\end{equation}
The formula (\ref{n0}) has a transparent physical meaning. Indeed,
$n_0$ is the most typical value of  the number of particles at an
active site during an avalanche, which ensures the density of
particles at the other sites of the lattice to be equal to
$\rho_{c}$.\ In other words, the avalanche dynamics in the system
above the critical point is self-organizing in a sense that it
maintains parameters of the medium outside the active site at the
critical point. In the vicinity of $n_0$, the function $P(n)$
shows a Gaussian-like form with the power law prefactor
$n^{\alpha-2}$. Replacing the sum (\ref{V}) by the integral, which
can be evaluated in the saddle point approximation, we obtain the leading part
of $V$, exponentially growing with $N$ for $\rho>\rho _{c}$
\begin{equation}
V\sim\left(  \rho-\rho_{c}\right)  ^{\alpha-1}N^{\alpha-1/2}a^{N}
\label{Vsupercit}%
\end{equation}
where $a=\left( \rho/{\rho_{c}}\right) ^{\rho}\left(  (1-\rho)/
({1-\rho_{c}})\right)  ^{1-\rho}$. Thus, the average avalanche size
either exponentially diverges with $N$ for $\rho$ above  $\rho_{c}$
or tends to a finite limit below  $\rho_{c}$.
These two regimes are separated by the point $\rho=\rho_{c}$,
where the power law growth of the avalanche size takes place
\begin{equation}
V_{c}  \sim N^{\alpha/2} \label{Vc}.
\end{equation}
The crossover from sub- \ to super-critical regime through the
critical point can be viewed in the function obtained from another
expansion of  Eq. (\ref{P(n)})  for $1 \ll n_{0}\ll N$, namely, $V \sim
V_{c} g_{\alpha}\left(  t\right)$, where
\begin{equation}
  g_{\alpha}\left(  t\right)  =
\frac{2e^{t^{2}}}{\Gamma\left(  \alpha/2\right)
}\int_{-t}^{\infty}dx\left(  x+t\right)  ^{\alpha-1}e^{-x^{2}}
\label{V1}
\end{equation}
is a function depending on the parameter $\alpha$ and  variable
$t=\sqrt{N\left(  1-\rho\right)  \rho/2}\ln \frac{\left(
1-\rho_{c}\right)  \rho}{\left(  1-\rho\right)  \rho_{c}}$, which
shows a distance from the critical point. Near critical point
$t\sim\left(\rho-\rho_{c}\right)\sqrt{N}$. The function
$g_{\alpha}\left(  t\right)  $ is equal to $1$ when $t=0$, and
decays as $|t|^{-\alpha}\ $ when $t\rightarrow-\infty$, thus
eliminating the dependence of $V$ on $N$ in the subcritical
regime.
Next, we consider particular examples of toppling probabilities
discussed in \cite{piph}.

\textit{In the integrable version of the} ASAP, the toppling
probabilities are defined by the recursion relations
$\mu_{2}\equiv\mu,$ $\mu_{n+1}=\mu (1-\mu_{n})$. In the notations
used, this case corresponds to $\mu_{\infty }=\mu/\left(
1+\mu\right)  $ and $\ \alpha=2$.  The sum in (\ref{V}) can be
reexpressed in terms of the infinite sum of hypergeometric
functions $\left. _{2}F_{1}\left(  a,b,c,d\right)  \right.  $
\begin{eqnarray}
V  &  =\frac{1+\mu}{\mu}\frac{N}{N-p+1}\sum_{k=0}^{\infty}\left[  -\left(
-\mu\right)  ^{k+1}\right]  \times   \label{velos2F1}\\
&  \left.  _{2}F_{1}\left(  2,1-p,2+N-p,\left(  -\mu\right)  ^{k+1}\right)
\right.  . \nonumber
\end{eqnarray}
This form is useful because all terms of the sum are of the same
structure. The terms with even $k$  look like the average avalanche size
of the ASAP with particular toppling probabilities,
$\mu_n=\mu^{k+1}$, independent on $n$, hence having its own
critical density $\rho_{c}^{[k]}=1/\left( 1-\left( -\mu\right)
^{k+1}\right) $, while the odd $k$-th terms are not singular in
all phase space. Therefore, for $\ \rho_{c}^{\left[ 2n\right]
}<\rho<\rho_{c}^{\left[ 2n+2\right] }$ the growing part of the
average avalanche size is given by the sum of $n+1$ exponentially growing
summands similar to one in Eq.(\ref{Vsupercit}) up to the change
of $\rho_c$ to $\rho_{c}^{[2k]},(k=0,2,\ldots,2n)$ and $\alpha
=2$. In the thermodynamic limit the term with $k=0$  is dominating.
 At  the critical point
$\rho_c=1/\left(  1+\mu\right)  ,$ the exact equality \cite{spirid}
\begin{equation}
_{2}F_{1}\left(  2,1-p,2+\mu p,-\mu\right)  =(1+\mu p)/({1+\mu} ),
\label{exact}
\end{equation}
leads to the surprising result,$V_c=N+O(1)$, which shows that the
growing part of the average avalanche size
 does not depend on $\mu$  in
the integrable case. The expansion below the critical point, $\rho<\rho_c$,
confirms the thermodynamics value and  $1/N$ correction obtained
from the Bethe ansatz solution \cite{piph,pph}.

\textit {For two-parametric} ASAP  \cite{piph} with toppling probabilities
defined by
${\mu_n}/({1-\mu_n})=2\mu(n-2+\alpha)/(({1-\mu})n\alpha)$,  we have
$\mu_{\infty}=2\mu/\left(  \alpha+2\mu-\alpha\mu\right)$ and 
\begin{equation}
\small
V=\frac{N}{N-p+1} \left. _{2}F_{1}\right. \left(
\alpha,1-p,2+N-p,\frac{-2\mu} { \alpha\left(  1-\mu\right) }
\right).
\end{equation}
 When $\alpha=2$, we have $\mu_n=\mu_\infty=\mu$, 
 $\rho_c=1-\mu$, and  $V_c=p$.

Let us consider  a connection of these results with MZ arguments \cite{MZ}.
Considering the form of transition probabilities
of Eq.(\ref{rw}) for $\rho=\rho_c$, one can see that the height of an active site
in the ASAP also performs the simple random walk when $\alpha=2$.
However, for  $\alpha\neq 2$ the normalized transition probabilities, defined as
$p(n \to n \pm 1)={w_{n,n\pm1}}/(w_{n,n+1}+w_{n,n-1})$,
contain the nonuniform bias term, $p(n \to n \pm 1)=1/2(1 \pm (\alpha-2)/(2n)+O(1/n^2))$,
 which corresponds to the process first considered by Gillis \cite{gillis}.
Depending on whether $\alpha$ is larger or less than two,  it can be either
positive or negative, respectively, enhancing or suppressing
the avalanche spreading.

As usual, we assume the following scaling ansatz for avalanche size and  time distribution
$P(s)=s^{-\tau}g(s/s_0)$ and $P(t)=t^{-\tau_t}g(t/t_0)$ in the vicinity of the critical point,
where $g(x)$ is a scaling function, and $s_0,t_0$ are time and size cutoffs.
The time cutoff $t_0$ also plays the role of correlation length.
Knowing asymptotics for mean conditional time of the first return for the Gillis
random walk \cite{hughes}, we can  obtain the
avalanche time critical exponent, $\tau_t=5/2-\alpha/2$ for $1<\alpha<3$.
To probe the avalanche size statistics
we should note that unlike the exponent $\tau_t $, the  dimension of avalanches $D$ does not depend
on $\alpha$ coinciding with unbiased case, which can be directly checked by multiplying the original equation
(\ref{rw}) by $n^2$ and  summing by parts. This yields $\langle n^2 \rangle \sim t$ and hence $D=3/2$ and
$\tau=2-\alpha/3$. The other critical exponents characterizing underlying interface dynamics
remain unchanged comparing to unbiased MZ case. Particularly, the
 exponent characterizing the correlation length below the critical point,
  defined in our case by  $t_0\sim (\rho_c-\rho)^{-\nu}$, can be uniquely fixed as  $\nu=2$ by
 requiring the obtained exponents to be consistent with the  average avalanche size
 below $\rho_c$, Eq.(\ref{Vsubcrit}).
In the critical point one gets from the Eq.(\ref{Vc}) $t_0\sim N$
 irrespectively of $\alpha$,
 i.e. the time cutoff grows as a system size, which determines the dynamical exponent $z=1$.
 We should note that although the correlation length demonstrates
  universal behavior independent on $\alpha$, both in the subcritical (KPZ) and critical (TI) regime, the results for the
mean avalanche size in the  intermediate regime (\ref{V1}) show the remaining  dependence on $\alpha$.
 The exponent $\theta=1$ just above the depinning threshold follows from the
Eq.(\ref{n0}), which gives a characteristic increase of the interface height after one
 avalanche passage proportional to $(\rho-\rho_c)$. The roughness exponent $\chi=1/2$
 is due to equiprobability of particle configurations established above.
We should note that the case $\alpha\neq 2$ violates the scaling relation 
\begin{equation}
\tau=1+(d-1/\nu)/D
\end{equation}
 valid for a wide class of the models  \cite{MZ} probably because
the interaction is not uniform with respect to the height of an avalanche. Rigorous explanation of this
violation is a matter of further analysis.

In summary, we have found that  the asymmetric avalanche process  (ASAP)  
\cite{piph}  incorporates
annealed KPZ and quenched TI interface dynamics for $\rho < \rho_c$ and
$\rho > \rho_c$, respectively.  All critical
exponents characterizing the interface behavior in both
classes are obtained exactly and shown to coincide with those known before.
Nevertheless, the exact calculation
of the average avalanche size and random walk arguments shows
that critical exponents of the avalanche distributions
are the continuous functions of the parameter $\alpha$ responsible for the
asymptotics of toppling probabilities and
the scaling relation between avalanche-size exponent $\tau$ and
correlation-length exponent  $\nu$   \cite{MZ} valid for many  other models
 is violated in the ASAP with  $\alpha \ne 2$. 

We are grateful to V.P. Spiridonov and Yan-Chr Tsai 
for useful discussion and the former for help in proof
of formula (\ref{exact}).
This work was supported in part by the National Science Council of the
Republic of China (Taiwan) under Grant No. NSC 91-2112-M-001-056 and Russian Foundation
for Basic Research under Grant No.03-01-00780.

\end{document}